\documentclass{article}
\usepackage{epubtk}    % calls hyperref, fancyvrb, and LRSP macros
\usepackage{epstopdf}
\usepackage{natbib}
\usepackage{booktabs}  % possibly also \usepackage{longtable}
\usepackage{graphicx}  % must come after epubtk

\begin{document}

\title{Spectropolarimetry of Atomic and Molecular Lines near 4135nm}

\author{\epubtkAuthorData{M Penn, H Uitenbroek, TA Clark, R Coulter, PR Goode, W Cao}
{National Solar Observatory\\
950 N Cherry Av, Tucson AZ 85726}
{mpenn@nso.edu}
{ http://www.noao.edu/noao/staff/mpenn/}
}

%\date{14 Oct 2014}
%\date{11 Nov 2014}
%\date{18 Nov 2014}
%\date{19 Nov 2014}
%\date{05 Dec 2014}
\date{12 Oct 2015}
\maketitle

\begin{abstract}
New spatially scanned spectropolarimetry sunspot observations are made of photospheric
atomic and molecular absorption lines near 4135nm.
The relative splittings among several atomic lines are measured
and shown to  agree with values calculated with configuration interaction and intermediate coupling.
Large splitting is seen in a line identified with Fe I at 4137nm,
showing multiple Stokes V components and an unusual linear polarization.
This line will be a sensitive probe of quiet Sun magnetic fields,
with a magnetic sensitivity of 2.5 times larger than that of the well-known 1565nm Fe I line.
\end{abstract}
\epubtkKeywords{Solar magnetic fields, Solar atmosphere, Detectors}
\newpage

%=================================================
%=================================================
\section{Introduction}
\label{sec:section1}

Measurements of solar magnetic fields are essential to modern research
efforts, and newly built telescopes and their instruments
(i.e. NST and CYRA, 
\cite{2010AN....331..636C})
and telescopes and instruments currently under construction
(i.e. DKIST and Cryo-NIRSP,
\cite{2003SPIE.4853..215L})
have new infrared magnetic observations in their core science programs.
Historically measurements of magnetic fields at infrared wavelengths have enabled
new scientific advances in solar physics, and these instruments intend to 
capitalize on such observations and seek new infrared tools as well.
It is well-known that for solar spectral lines with identical magnetic splitting,
the relative sensitivity of lines to solar magnetic fields
increases with increasing wavelength.
As a method to determine which spectral lines to target for increased
accuracy in measuring the solar magnetic field,
the concept of the magnetic resolution of a spectral line
can be used.
The magnetic sensitivity is determined by the ratio
of the magnetic Zeeman splitting
$ \lambda^2 g_{eff}$
and the Doppler line broadening
$\lambda $
and thus varies as
$\lambda g_{eff}$.
Currently the most sensitive probes of the solar magnetic field
include the Fe I 1564.8nm spectral line
($\lambda g_{eff} = 4695$)
for photospheric fields,
the Ti I 2233nm line 
($\lambda g_{eff} = 5778$)
for fields in cold sunspot umbrae,
and the Mg I 12381nm line 
($\lambda g_{eff} = 12318$)
for penumbral and plage magnetic fields.
Reports of other very sensitive spectral lines 
near 4135nm have been
made but no measurements have been published.

Identification of infrared lines is still a challenging task,
and several lines
remain unidentified even today
\citep{1995ASPC...81...88G}.
From laboratory work, lines from high-energy transitions in Fe I,
in particular from the
$3d^64s6s$
to 
$3d^64s6p$
configurations have
"...fragmented analysis with only a few lines known."
\citep{1988JOSAB...5.2264J}
For lines in the 4135nm region, only one identification is given by
\cite{1994ApJS...94..221N}
for the 4136.4908 nm line.
The energy levels for the transition correspond to 
56541.592 
$cm^{-1}$
and 
54124.740
$cm^{-1}$
for the upper and lower levels, respectively,
producing a transition at 
2416.86
$cm^{-1}$
or 
4136.48579nm.
The electronic configurations and terms listed by 
\cite{1994ApJS...94..221N}
are
$3d^64s(^6D)6p(^7P^0)$
and
$3d^64s(^6D)6s(g^7D)$
for the upper and lower levels.
Using L-S coupling 
(e.g. \cite{1969tzm..book.....B})
to determine the magnetic sensitivity of
this transition gives
$g_{eff} = 1.70$.

Using theoretical models and fitting to laboratory spectra,
\cite{2011CaJPh..89..417K}
has published a set of infrared line identifications with
electron configurations and terms
for the upper and lower levels for these lines,
as well as the Lande-g
values for the levels for the lines from Fe I
\citep{2011CaJPh..89..417K}.
These calculations differ from
previous work
in that the energy levels are not specifically assigned
to an individual electronic configuration and term,
but rather to a superposition of overlapping 
configurations.
These configuration interactions
are solved to produce an
eigenvector of electronic states
with a set of contribution coefficients for each state.
The energy level is then composed of the sum 
of these individual levels.
With a mixture of electronic configurations for the upper and
lower states of
the 4136.5nm transition,
LS coupling is applied to each possible combination
of upper and lower states
weighted with the proper coefficients to 
produce a
value for the magnetic
sensitivity of the spectral line;
this calculation is called 
intermediate coupling.
For the Fe~I 4136.5nm line
the computed value for intermediate coupling is
$g_{eff} = 1.62$.
While intermediate coupling gives a similar value to LS coupling in this example,
this is not always the case.

To determine the relative importance of mixed
terms for the energy levels which produce a particular spectral line,
the Zeeman splittings of spectral lines
measured in the laboratory can be compared 
against
the predictions from
both L-S
and intermediate
coupling
\citep{Curtis2003book}.
The observations presented in this work provide the first such
experimental data from the Sun which can be used to
address such questions about
several spectral lines near 4135nm.
Furthermore, the polarized spectral line profiles observed
in sunspots and presented in this paper will provide an
experimental constraint on the intermediate coupling
theory used to compute the Lande-g values.
While the values of 
$g_{eff}$
have been computed for intermediate coupling, 
the polarized spectral line profiles themselves have not been computed.
These profiles are critical for using these spectral lines to
infer the magnetic fields on the Sun.
They can be computed by first computing the expected LS coupling
Zeeman patterns among all the configuration interaction levels, 
and then weighting those patterns with the proper coefficients.

In section 2 we review the atomic levels the relevant
atomic and molecular lines
near 4135nm,
in section 3 we discuss new solar observations of
this part of the spectrum,
and then in section 4 we present
the wavelengths and magnetic splittings measured for these lines.

%=================================================
%=================================================
\section{Lines near 4135nm}

Observations of the quiet Sun photospheric spectrum 
near 4135nm were made
from space during the ATMOS experiment
\citep{1989hra1.book.....F}
and 
from the ground at the National Solar Observatory's
McMath/Pierce Solar Facility (McM/P) 
\citep{2003assi.book.....W}.
Observations of the sunspot umbral spectrum near 4135nm were
also made at the McM/P
( \cite{2002sus..book.....W} and
\cite{1992adsu.book.....W}).
Line identifications are given in the McM/P spectra,
and have been made for the ATMOS spectra too
\citep{Geller1992.ATMOS.Lines}.

Table~1 
lists the observed atomic spectral lines in this part of the 
solar spectrum as seen in the quiet Sun photosphere.
Here the first wavelength values
(with the K subscript)
are taken from 
\cite{2011CaJPh..89..417K},
as are the values for
$log(gf)$.
The second wavelength values 
and the line depth values
(with the G subscript)
are taken from 
\cite{Geller1992.ATMOS.Lines},
and here the line depths have been converted to
continuum percentages.
Finally, the last two columns
(with the NAC subscript)
represent recent observations taken from the McM/P and
will be described in more detail in the following section.

\begin{table}[h]
  \caption{Atomic Spectral lines}
  \label{tab:table01}
  \centering
  \begin{tabular}{ccccccc}
    \toprule
     Species & $\lambda_K$ & log(gf) & $\lambda_G$ & Depth$_G$ & $\lambda_{NAC}$ &  Depth$_{NAC}$  \\
      & [nm] &  & [nm] &  & [nm] &  \\
    \midrule
     Si I & 4122.8421 & -2.190 & 4122.79 &  3.2 & 4122.90 & 0.6 \\
     Si I & 4132.7169 &  0.250 & 4132.72 & 16.0 & 4132.75 & 5.2 \\
     Fe I & 4136.4857 &  0.512 & 4136.47 &  7.9 & 4136.50 & 2.7 \\
     Fe I & 4137.0095 & -1.630 & 4136.97 &  3.1 & 4136.99 & 0.8 \\
     Fe I & 4139.2294 &  0.448 & 4139.18 &  6.8 & 4139.19 & 2.6 \\
     Si I & 4142.5516 &  0.580 & 4142.47 &  9.6 & 4142.48 & 5.2 \\
  \end{tabular}
\end{table}

The infrared solar spectrum has few atomic lines,
and the lines that are present are often weak.
In order to produce lines at these wavelengths,
the energy difference between the upper and lower levels must
be small.
The energy difference between levels near the ground state of most
atoms is larger, resulting in lines in the UV or visible.
Only as one moves to higher excited states does one find upper 
and lower levels with relatively close spacing in energy which
are able to produce spectral features in the infrared.
For both Fe I and Si I, the upper and lower energy levels
for these lines are rather close to 
the ionization energies,
which are about
63737
cm$^{-1}$
for Fe I 
\citep{1985aeli.book.....S}
and 
65748
cm$^{-1}$
for Si I
\citep{1983MartinZalubas}.
As the temperature of the solar plasma drops from the
value in the quiet Sun
to lower values in sunspots,
the electron populations in these
energy levels are much smaller.
For this reason the strength of these absorption lines
is greatly reduced in sunspots,
and these spectral lines are not useful tools for examining the
physical conditions in sunspot umbrae.
In the penumbrae of sunspots,
it is expected that the lines are probing the physical
conditions in the hotter plasma, or brighter penumbral structures.
It is expected that these spectral lines will prove most useful as
diagnostics for the quiet Sun plasma.

Table~2 lists the energy levels involved with each of these
lines,
and the primary level configurations for each level
(although it is important to remember that other configurations
are important under the condition of configuration interaction).
The identifications are taken from
\cite{Geller1992.ATMOS.Lines}
and
\cite{2003assi.book.....W},
with the exception of the 4137nm line, 
which is here associated with an Fe~I transition listed by
\cite{2011CaJPh..89..417K}.
Electron configuration and terms
for Si I
were taken from 
\cite{1983MartinZalubas}
and they report mixing percentages for configuration interactions
in the energy levels
at
54871, 56690, and 58893
cm$^{-1}$.
For Fe I the configurations and terms were taken from 
\cite{1994ApJS...94..221N}.
For these primary configurations, the values for
the L,S and J quantum numbers are listed,
along with the Lande
g value
(from
\cite{2011CaJPh..89..417K})
for the energy levels.

\begin{table}[h]
  \caption{Line Identifications}
  \label{tab:table02}
  \centering
  \begin{tabular}{ccclc}
    \toprule
     Species & $\lambda_K$ & Level Energy & Level Configuration & Level g$_K$  \\
      & [nm] &  [cm$^{-1}$] & (primary)  &  \\
    \midrule
     Si I & 4122.8421 & 54871.031 & $3s^23p5s~^1P^0$ &  \\
          &           & 57295.881 & $3s^23p5p~^3P$   &  \\
	& & & & \\
     Si I & 4132.7169 & 56690.903 & $3s^23p4d~^3P^0$ &  \\
          &          & 59109.959 & $3s^23p(^2P^0_{3/2})4f~^2[5/2]$ &  \\
	& & & & \\
     Fe I & 4136.4857 & 54124.740 & $3d^64s(^6D)6s~g^7D$ &  1.650  \\
          &           & 56541.592 & $3d^64s(^6D)6p~^7P^0$  &  1.587  \\
	& & & & \\
     Fe I & 4137.0095 & 54747.594 & $3d^64s(^6D)6s~g^7D$ & 2.999 \\
          &           & 57164.140 & $3d^64s(^6D)6p~^7D^0$ & 2.642 \\
	& & & & \\
     Fe I & 4139.2294 & 54611.706 & $3d^64s(^6D)6s~g^7D$ & 1.997 \\
          &           & 57026.956 & $3d^64s(^6D)6p~^7F^0$ & 1.659 \\
	& & & & \\
     Si I & 4142.516 & 58893.400 & $3s^23p4d~^1F^0$ &  \\
          &          & 61306.713 & $3s^23p(^2P^0_{1/2})5f~^2[7/2]$ &   \\
	& & & & \\
  \end{tabular}
\end{table}

For the case of LS coupling, the Land\'{e} g value for the energy level
can be given with the well-known expression
(e.g.  \cite{1969tzm..book.....B} equation 1)
and then the value for
$g_{eff}$
for the spectral line can be computed from the values for the upper
and lower levels as:
(\cite{1969tzm..book.....B} equation 6):
$g_{eff} = \frac {1}{8} {(J_u-J_l)(J_u+J_l+1.0)(g_u^2-g_l^2)}$.

The molecular lines in this region of the spectrum are 
dominated by OH and SiO lines.
\cite{Geller1992.ATMOS.Lines}
lists several OH lines 
in this part of the spectrum as observed 
from the ATMOS data at
wavelengths of
4133.26, 4135.10, 4136.79 and 4142.69nm.
In this spectral region,
\cite{2002sus..book.....W}
identify about 30 lines 
from SiO, and three lines from CO.
They also identify 7 OH lines,
including all of the OH lines from
\cite{Geller1992.ATMOS.Lines}
plus new identifications at roughly
4136.5 , 4137.9
and 
4139.2nm.

%=================================================
%=================================================
\section{Observations and Data Analysis}

Two sets of observations have been used to investigate
these spectral lines,
both taken at the 1.6m diameter McM/P main telescope
and main spectrograph
\citep{1964ApOpt...3.1337P},
but using two different instruments.
The first measurements were taken
with the NIM instrument
\citep{1992AAS...181.8101R}
and measured the intensity profiles across several sunspots
in
the late 1990's.
The more recent measurements were taken using the
the NSO Array Camera (NAC)
with a 1024 x 1024 InSb Alladin 3 array as the detector.
While the NIM observations include several additional lines 
of interest, including lines at 4056 and 4122.8nm,
and while the NIM observations were the inspiration for the recent NAC observations, 
here only the NAC observations will be discussed,
and in future work
more details of the NIM observations will be presented.

In the NAC observations,
the wavelength selection in the dewar was done using a 1-inch round cold filter
(at roughly 60K) with a 18 nm bandpass centered at 4137 nm
with a peak transmission of 75 \%.
Since the main spectrograph is warm,
the observations contain a large background level.
Exposure times were
25 microseconds per frame.
The McMath-Pierce optics are expected to
produce a very small instrumental polarization at this wavelength,
and the polarization crosstalk cannot be measured with this data.

The polarization analysis for the NAC observations
was done using optics positioned at the spectrograph exit port.
A reimaging bench using two CaF2 lenses collimated the spectrograph exit
and reimaged it onto the NAC detector. 
A rotating waveplate was positioned at the spectrograph exit focal plane,
and then a linear polarizer was used in the collimated beam.
Since the rotating waveplate is positioned near an image plane, 
deflection of the beam is not an issue, 
but dust and other transmission features on the waveplate 
are challenging to remove during calibration.
A slow-chopping method was used by rotating the waveplate between
exposures.
For sunspot scans,
a scanning mirror stepped the solar image perpendicular to
the spectrograph slit at the end of a sequence of polarization exposures.
Background and flat field corrections were made for each exposure at
the waveplate positions.
The waveplate rotations positions were calibrated using
a second linear polarizer in the beam.

Table~3 lists details of both the NIM and the NAC observations.

\begin{table}[h]
  \caption{McM/P Observations}
  \label{tab:table03}
  \centering
  \begin{tabular}{llllll}
    \toprule
    Number & Date & Time & Wavelength & Sunspot & Stokes type \\
      & & [UT] & [nm] & & \\
    \midrule
% 20130502 spot11732.00001 - 00250
    1 & 2013/05/02 & 21:08 & 4120-4145 & NOAO 11777 &I \\
% 20130905 spot05
    2 & 2013/09/05 & 21:10 & 4136-4139 & NOAO 11836 &I,V \\
% 20130905 spot06
    3 & & 21:22 & 4132-4136 & NOAO 11837 &I,V \\
% 20130905 spot07
    4 & & 21:29 & 4136-4139 & NOAO 11838 &I,V \\
% 20130912 spot_seq.00011.fit
    5 & 2013/09/12 & 17:07 & 4136-4139 & NOAO 11841 &I,Q,U,V \\
% 20130925 spot2.01247
    6 & 2013/09/25 & 18:31 & 4136-4139 & NOAO 11846 &I,Q,U,V \\
  \end{tabular}
\end{table}

Because the McM/P adaptive optics system was not used for these
observations,
the solar pointing of different waveplate sequences is not stable.
Only exposures where two or more spectral lines are simultaneously measured
are used to compute the relative splitting ratios between lines, 
in this way the magnetic fields sampled by the different lines is identical and
the different splitting represent the inherently different responses
from the spectral lines. 
The NAC exposure times and the waveplate rotation were run in open-loop mode,
resulting in some overhead time where photons were not collected.

%=================================================
\section{The Intensity and Polarization Spectra}

\subsection{Intensity Spectra 4120-4145nm}

The four strongest lines in this spectral region
are two each from Fe I and Si I.
There are also two weaker lines, one each from Fe I and Si I.
There are also several telluric absorption lines.
The spectra are shown in Figure~1.
The averaged NAC spectrum has about 10 times better signal to noise
than in the spectrum from 
\cite{2003assi.book.....W}
but the lines appear weaker in the NAC spectrum due to the
reduced spectral rsolution compared to
\cite{2003assi.book.....W}.
Many frames were stitched together to produce this spectrum,
where each single NAC image
covers about 4nm of the solar spectrum.

The wavelength calibration of this data was done using a set of six
telluric absorption lines located at
4131.79, 4133.15, 4134.69, 4137.94, 4141.03,
and 
4144.12nm as measured in the
\cite{2003assi.book.....W}
No corrections were made for the Doppler shift caused by the 
relative motion of the Sun and the telescope,
nor for the gravitational redshift of the Sun.
As shown in Table~1,
the three Si I line centers were measured at positions of
4122.90,
4132.75
and
4142.48nm
and the three Fe I line centers were measured to be at
4136.50,
4136.99
and
4139.19nm.
The uncertainty in these observed wavelengths is about 
$\pm$0.005nm.

%\epubtkImage{Fig01.png}{%
%  \begin{figure}[htbp]
%    \centerline{\includegraphics[width=0.9\textwidth]{figure01.eps}}
%    \centerline{\includegraphics{Fig01.png}}
%    \caption{Intensity spectrum from 2013 May 02
%             data produced by blah frames taken with the NAC at the McM/P,
%             plotted with the FTS atlas spectrum from the NSO archives.
%             While the NAC spectrum shows weaker lines likely due
%             to worse spectral resolution or improperly removed background,
%             the signal to noise level is about 10 times better than in
%             the FTS spectrum because of the large number
%             of pixels which were averaged.}
%    \label{fig:fig01}
%\end{figure}}

The observed line depths from the NAC observations correlates
well with the line depths from ATMOS
\citep{Geller1992.ATMOS.Lines}
with the exception of the 4142nm Si I line.
The NAC line depths are roughly one-third of the ATMOS line depths,
except that in the NAC observations the 4142nm line is
about twice as deep as this relationship would suggest.

Figure~2 shows a spectral frame from the 25 Sep 2013 observations;
in this frame, the upper and lower sections of the slit
show the quiet Sun spectrum,
the middle of the slit crosses a sunspot umbra and
the other parts of the slit show signal from the penumbral regions.
The spectral range covers from about 4136 to 4141nm.
This frame was produced by averaging the spectrum from 
240 individual exposures, after each was shifted to align the
sunspot position along the slit.
Residual seeing motions perpendicular to the slit were not 
corrected, and so some spatial smearing is introduced.
The umbral regions of this sunspot show
a continuum brightness of about
0.72 times the quiet Sun brightness.
As shown in Figure~2, there are several spectral lines 
which are not present in the quiet Sun regions,
but appear in the sunspot umbral regions.
They are identified using
\cite{2002sus..book.....W}.
The strongest molecular lines seen in the intensity spectrum 
are at the following measured wavelengths:
4136.48 (OH or SiO),
4136.85 (SiO),
4137.48 (SiO or CO),
4137.98 (OH),
4138.54 (SiO),
4139.40 (OH),
4139.63 (SiO),
and
4140.75 (SiO).
In all cases the error in the line center wavelengths is
estimated to be about
$\pm$0.02nm.
The only ATMOS molecular identifications in this range listed in 
\cite{Geller1992.ATMOS.Lines}
are from OH at 4136.79 and 4139.26nm
and these lines are not seen in the sunspot spectra
of
\cite{2002sus..book.....W},
nor in these NAC sunspot spectra.

%\epubtkImage{QSunspec_20130502.png}{%
  %\begin{figure}[htbp]
    %\centerline{\includegraphics[width=0.9\textwidth]{Figure2.eps}}
    %\caption{Stokes V spectrum from 23 Sep 2013
             %data produced by blah frames taken with the NAC at the McM/P,
             %plotted with the FTS atlas spectrum from the NSO archives.
             %While the NAC spectrum shows weaker lines likely due
             %to worse spectral resolution or improperly removed background,
             %the signal to noise level is about 10 times better than in
             %the FTS spectrum because of the large number
             %of pixels which were averaged.}
    %\label{fig:figure03}
%\end{figure}}

%=================================================
%=================================================
\subsection{The Polarization Spectrum 4135-4139nm}

The Stokes spectra of these lines show a wealth of detail.
In Figure~3, a Stokes~V spectral frame is shown;
this corresponds to the same position as the Stokes~I
frame shown in Figure~2.
This and subsequent Stokes~Q and Stokes~U frames are
made by analyzing 30 waveplate rotations,
each of which was sampled with 8 exposures.
As in the Stokes~I spectral frame,
motion of the sunspot along the slit was corrected,
but motion perpendicular to the slit will add spatial smearing,
and possibly spectral broadening as slightly different 
magnetic fields are sampled in each exposure.
It is important to note that all three spectral lines are
impacted in identical ways,
since they all appear on the same spectral frame.
This frame prominently shows the circular polarization
signature from the three Fe~I spectral lines,
as well as the spectra from four molecular lines confined
to the sunspot umbra.

The averaged spectral frame shown in Figure~3 reveals that the McM/P observations
from the NAC currently lack the signal to noise needed to measure the
magnetic fields in the quiet Sun outside of the penumbra.
The spectral lines sample the penumbral fields well,
but when the field strength increases some of the 
Zeeman components from 4136 and 4137 blend.
The lines are very weak in the sunspot umbra, and do not seem useful
for measuring the magnetic fields there.

%\epubtkImage{QSunspec_20130502.png}{%
  %\begin{figure}[htbp]
    %\centerline{\includegraphics[width=0.9\textwidth]{Figure3.eps}}
    %\caption{Stokes V spectrum from 23 Sep 2013
             %data produced by blah frames taken with the NAC at the McM/P,
             %plotted with the FTS atlas spectrum from the NSO archives.
             %While the NAC spectrum shows weaker lines likely due
             %to worse spectral resolution or improperly removed background,
             %the signal to noise level is about 10 times better than in
             %the FTS spectrum because of the large number
             %of pixels which were averaged.}
    %\label{fig:figure03}
%\end{figure}}

In order to examine the relative magnetic splittings, 
the NAC Stokes~V profiles were examined using a number of techniques.
The analysis was restricted to only spectral frames showing two or more lines;
in this way the spectral lines are identically affected by seeing conditions,
and the spectral lines sample the identical solar magnetic fields.
The splitting was examined by eye using spectral profiles,
and then the Stokes signals in several frames were examined by hand.
Another more detailed analysis involved
finding the wavelength positions of the Stokes~V
$\sigma$
spectral components at several slit positions,
and fitting a polynomial function to these splittings to
derive the splittings along the whole slit.
In most cases both 
$\sigma$
components were used for these measurements,
however when the components were blended or outside of
the spectral frame,
one component was measured relative to the line center position.

The measured splittings for all the lines were then compared
to the splitting of the Fe~I 4139nm splitting.
The results are presented in Table~4.
The sources of error to consider in these measurements are the 
uncertainty in the wavelength position of the 
$\sigma$
component
(small)
the spatial scatter in the ratio of the splitting
in well-measured penumbral regions
(reported in Table~4),
and the uncertainty in the line center position in
cases where only one 
$\sigma$
component is measureable
(difficult to quantify, but estimated to be small).

From
\cite{1994ASSL..189.....S}
the Zeeman splitting of a line
$\Delta \lambda = 4.67 10^{-13} g_{eff} \lambda^2 B$,
where
$B$
is the solar magnetic field causing the splitting.
Here we ignore the 0.2\% changes due to
the different wavelengths of the lines and set
$\lambda$ for each line equal.
As long as the magnetic field being measured is identical
(a condition which is satisfied if the lines are in the same
spectral image)
then we can compute the ratio of the 
$g_{eff}$ 
values for each line as the ratio of the observed splittings
$\frac {g_{4136}} {g_{4139}} = \frac {\Delta \lambda_{4136}} {\Delta \lambda_{4139}}$.

The splitting for the Fe~I 4137nm line is a special case,
since as revealed by Figure~3,
the line seems to display multiple 
$\sigma$
components in the Stokes~V spectrum.
The individual components are split by factors of 
1.44 $\pm$ 0.05
and
2.55 $\pm$ 0.23
respectively.
The amplitude ratio of the two components is difficult to measure,
but is estimated at
1.75 $\pm$ 0.25,
with the component with the larger shift having the weaker
Stokes~V amplitude.
Using the averaging technique described by 
\cite{1969tzm..book.....B}
the 
Lande factor for this line is then determined to be
g$_{eff}$= 1.84$\pm$0.21.

\begin{table}[h]
  \caption{Zeeman Line Splittings Relative to Fe I 4139nm}
  \label{tab:table02}
  \centering
  \begin{tabular}{lllll}
    \toprule
     Atom & Wavelength & Kurucz Ratio &  Measured Ratio \\
      & [nm] & (vs 4139nm) & (vs 4139nm) \\
    \midrule
    Si I & 4132.7 &   & 1.00$\pm$0.12(3) \\
    Fe I & 4136.5 & 1.23 & 1.12$\pm$0.67 (2) \\
     &  &   & 1.28$\pm$0.14 (4) \\
     &  &   & 1.39$\pm$0.07 (5) \\
     &  &   & 1.32$\pm$0.03 (6) \\
    Fe I & 4137.0 &  2.13 & 1.84$\pm$0.21 (6) \\
%    Fe I & 4137.0 &  2.13 & 1.44$\pm$0.05 (6) \\
%    & &  & 2.55$\pm$0.23 (6) \\
  \end{tabular}
\end{table}

Four of the molecular lines seen in the sunspot from 20130925
show strong Stokes~V profiles.
These are the lines at:
4136.48 (OH or SiO)
4136.85 (SiO)
4137.98 (OH)
and
4139.40nm (OH).
The lines at 
4136.48
and
4137.98nm
show a Stokes~V profiles which has the opposite sense of
the atomic and the other molecular lines: 
these transitions have a negative
g$_{eff}$.

None of the molecular lines show completely resolved splitting profiles
in the sunspot umbra, rather they resemble the types of unresolved
Stokes~V profiles seen in weak magnetic fields.
Using the weak field approximation,
we can compare the relative 
g$_{eff}$
of the lines by examining the amplitudes of the 
Stokes~I  derivatives and Stokes~V profiles
of the lines.
From
\cite{1994ASSL..189.....S}
when the Zeeman splitting of a line is weaker
than the Doppler broadening, 
the Stokes~V signal can be approximated by
$V(\lambda) \approx B g_{eff} \frac {\partial I} {\partial \lambda}$
where
$B$
is the solar magnetic field causing the splitting.
Again, as long as the magnetic field being measured is identical
(a condition which is satisfied if the lines are in the same
spectral image)
then we can compute the ratio of the 
$g_{eff}$ 
values for line 1 compared to line 2 as the ratio of:
$\frac {g_{1}} {g_{2}} = \frac {V_1} {V_2} \frac {\partial I_2 / \partial \lambda} {\partial I_1 / \partial \lambda}$.
In Table~5 we list the ratios of the Stokes~I derivatives and V amplitudes 
of the lines,
normalizing all of the lines by the ratio
of the 
OH line at 4139.63nm.
While the Stokes~I derivatives for these lines are within a few percent of each other,
there are larger differences in the Stokes~V amplitudes.
The measurement errors in these relative values are small,
but the systematic errors caused by an uncertain background
subtraction are estimated at the level of 10\%.
In the two cases where the molecules show an inverted Stokes~V
profile, the ratio is listed as negative.
From the values in the Table it is clear that
the SiO line at 4136.85nm
has the largest value for 
g$_{eff}$
of all four of these lines.

\begin{table}[h]
  \caption{Stokes~V Amplitudes Relative to OH 4139.63nm}
  \label{tab:table03}
  \centering
  \begin{tabular}{lllll}
    \toprule
     Molecule & Wavelength & ${\partial I / \partial \lambda}$  & Stokes~V & g$_{eff}$ ratio\\
      & [nm] &  (vs 4139.63nm) & (vs 4139.63nm) & (vs 4139.63nm)\\
    \midrule
    OH (or SiO) & 4136.48 & 1.04 & -0.8 & -0.77 \\
     SiO &  4136.85 &  0.98 & 1.25 & 1.28 \\
     OH &  4137.98 & 1.05 & -0.8 & -0.77 \\
  \end{tabular}
\end{table}

The Stokes~Q and U polarization profiles are very strange for
the 
Fe~I 4137nm spectral line.
Both the 4136nm and 4139nm Fe~I lines show
typical linear polarization profiles with central 
$\pi$
components and magnetically shifted 
$\sigma$
components.
For both lines the 
$\sigma$
components have the same polarity at a given spatial location
in the penumbra.
However, the 
4137nm
line is different.
First, there is no unshifted central
$\pi$
component in the Stokes~Q or Stokes~U profile.
Secondly,
the magnetically split 
$\sigma$
components are visible, 
but they have the opposite polarity as the
$\sigma$
components for the 4136 and 4139nm lines.
This is unexpected, especially since the Stokes~V
profiles for all three Fe~I lines show the same 
polarity for the magnetically shifted
$\sigma$
components.
The molecular absorption lines in the sunspot umbra seem to show no
signals in the linear polarization Stokes~Q and Stokes~U data.

%\epubtkImage{QSunspec_20130502.png}{%
  %\begin{figure}[htbp]
    %\centerline{\includegraphics[width=0.9\textwidth]{figure4u.eps}}
    %\caption{Stokes U spectrum for 4137 and 4136nm Fe I line.
             %blah blah blah blah blah.
             %}
    %\label{fig:fig02}
%\end{figure}}
%
%\epubtkImage{QSunspec_20130502.png}{%
  %\begin{figure}[htbp]
    %\centerline{\includegraphics[width=0.9\textwidth]{figure4q.eps}}
    %\caption{Stokes Q spectrum for 4137 and 4136nm Fe I line.
             %blah blah blah blah blah.
             %}
    %\label{fig:fig03}
%\end{figure}}
%

%=================================================
%=================================================
\section{Conclusions}

The identifications of these infrared solar spectral lines with
particular transitions in Fe I and Si I is difficult,
but the latest observations from the NSO McM/P using the NAC
confirms several identifications made by Kurucz.
Relative to the Fe I line at 4139nm,
the Zeeman splitting of the 4136nm line is
$1.28\pm0.17$,
consistent with the prediction from the intermediate coupling model.
The Zeeman splitting of the Fe I line at 4137nm
shows two components,
with an intensity averaged splitting about 2 times
the 4139nm line splitting; 
this is roughly consistent with the value predicted
from the intermediate coupling.
The magnetic sensitivity of this line is therefore
very high, 
with a
$\lambda g_{eff} = 11600$.
While these lines may be useful for sunspot penumbra observations,
their main advantage will be realized with future observations with lower backgrounds
where they will be a critically useful 
diagnostic of quiet Sun magnetic fields.

%=================================================
%=================================================
\section{Acknowledgements}
\label{sec:acknowledgements}
Thanks to Robert Kurucz for clarifying discussions.
%=================================================

\newpage

\bibliography{penn_20151012.bib}

\epubtkImage{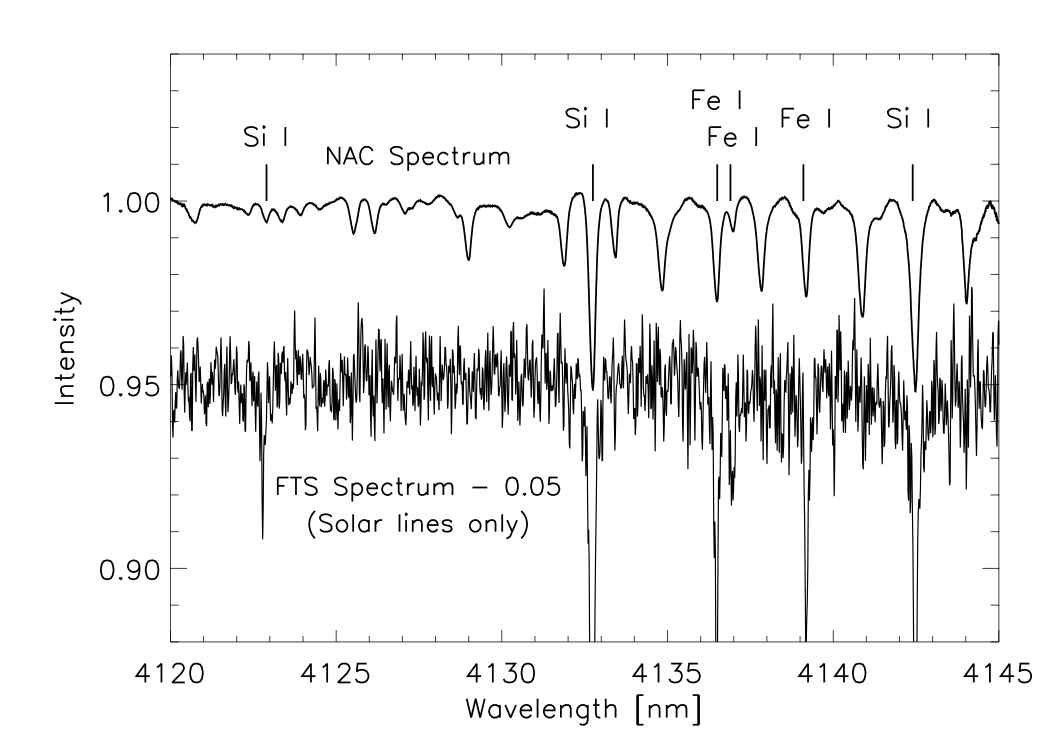}{%
\begin{figure}[htb]
  \centerline{\includegraphics[width=1\textwidth]{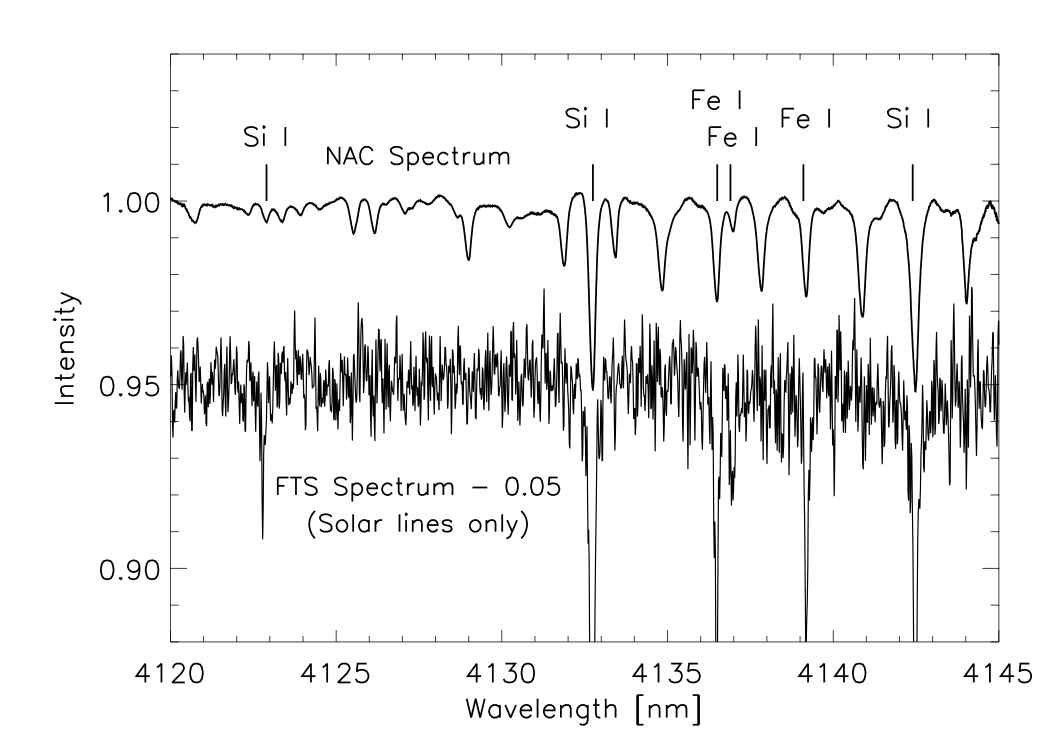}}
  \caption{Intensity spectrum near 4135nm.
The top plot represents the NAC spectrum, the bottom one is the FTS
spectrum, offset by a small amount.
}
  \label{fig:fig1}
\end{figure}}

\epubtkImage{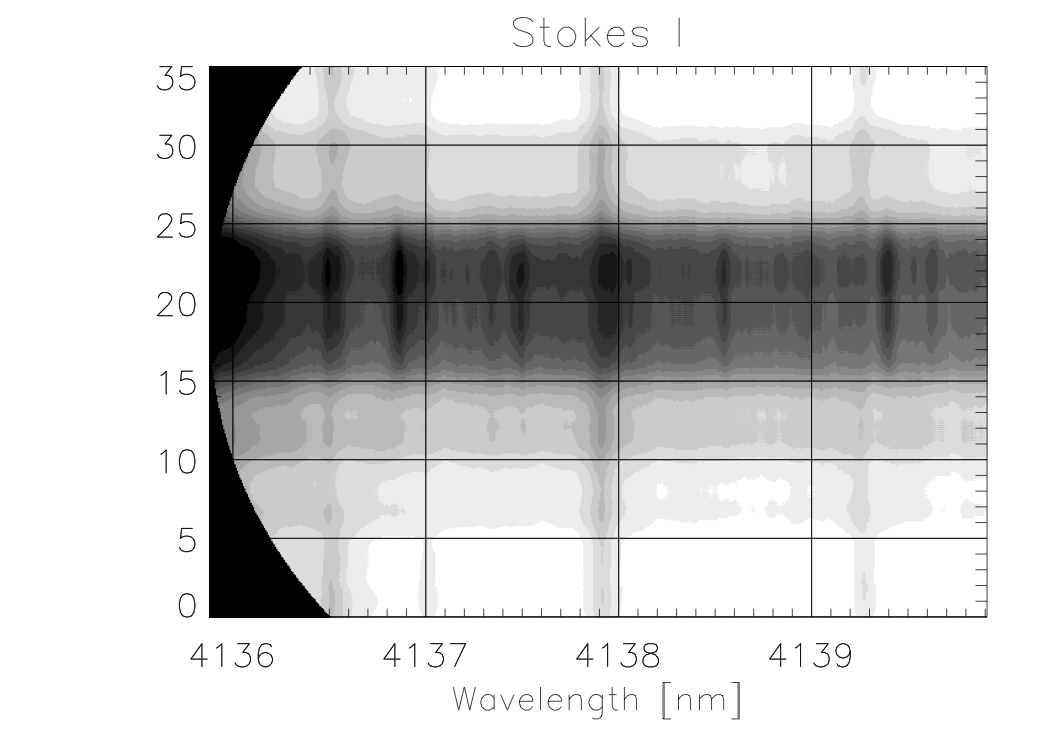}{%
\begin{figure}[htb]
  \centerline{\includegraphics[width=1\textwidth]{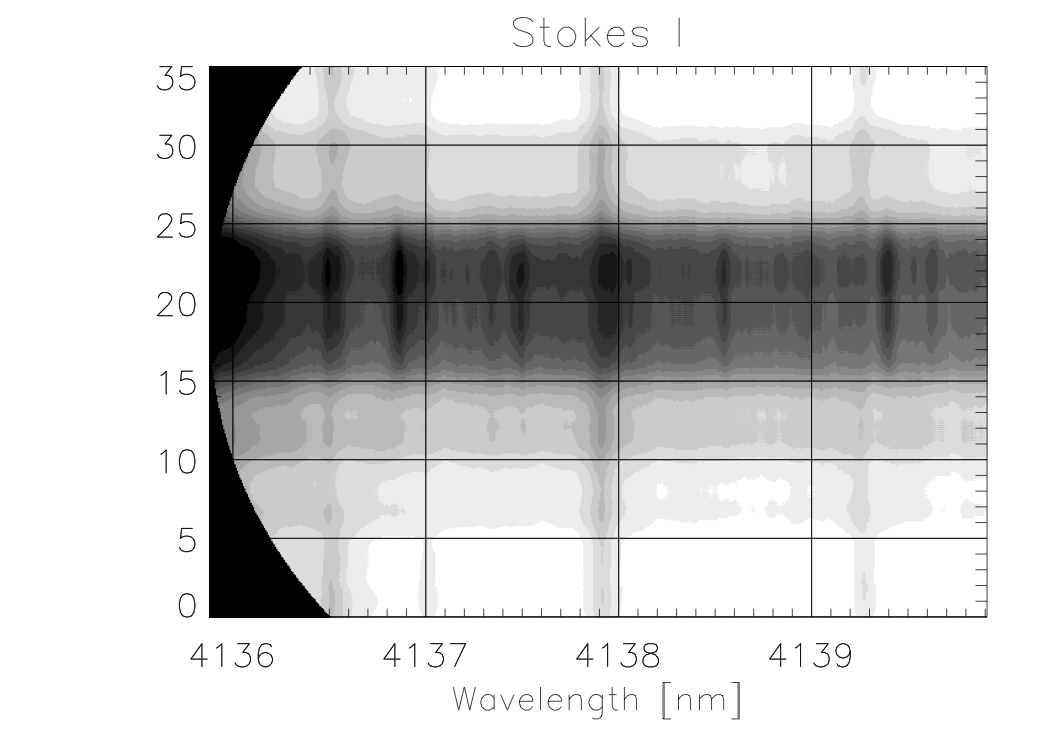}}
  \caption{Intensity spectral frame from sunspot observations near 4135nm.  
}
  \label{fig:fig2}
\end{figure}}

\epubtkImage{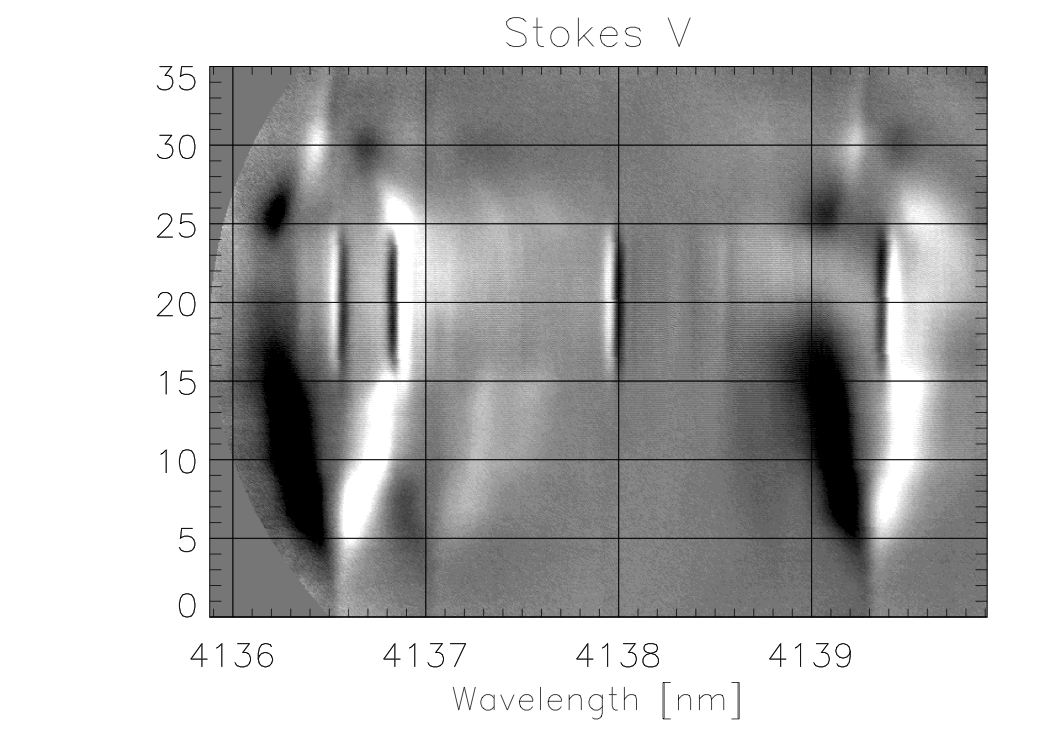}{%
\begin{figure}[htb]
  \centerline{\includegraphics[width=1\textwidth]{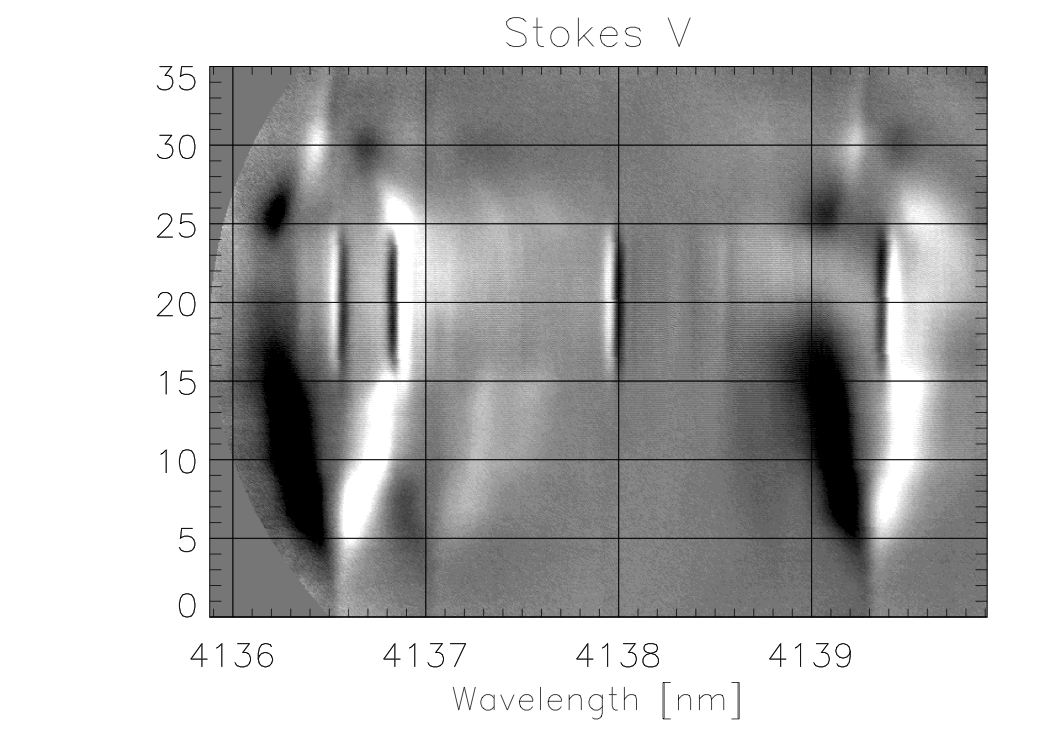}}
  \caption{Stokes V spectral frame from sunspot observations near 4135nm.
}
  \label{fig:fig3}
\end{figure}}

\end{document}